\documentclass[fleqn,usenatbib]{mnras}

\usepackage{newtxtext,newtxmath}

\usepackage[T1]{fontenc}

\DeclareRobustCommand{\VAN}[3]{#2}
\let\VANthebibliography\thebibliography
\def\thebibliography{\DeclareRobustCommand{\VAN}[3]{##3}\VANthebibliography}

\usepackage{graphicx}	
\usepackage{amsmath}	

\usepackage{microtype} 

\usepackage{etoolbox, siunitx}
\robustify\bfseries

\usepackage{booktabs} 

\usepackage[normalem]{ulem} 

\DeclareSIUnit\au{AU}
\DeclareSIUnit\Rsun{R_\odot}
\DeclareSIUnit\Rjup{R_\text{Jup}}
\DeclareSIUnit\Msun{M_\odot}
\DeclareSIUnit\Mjup{M_\text{Jup}}
\DeclareSIUnit\gyr{Gyr}
\DeclareSIUnit\ppt{ppt}
\DeclareSIUnit\ppm{ppm}

\title[Multi-Planet Circumbinary Migration]{Sculpting the circumbinary planet size distribution through resonant interactions with companion planets}

\author[Fitzmaurice, Martin \& Fabrycky]{%
        Evan Fitzmaurice$^{1}$,
        David V. Martin$^{1,2}$ \&
        Daniel C. Fabrycky$^{3}$
\\
$^{1}$Department of Astronomy, The Ohio State University, 4055 McPherson Laboratory, Columbus, OH 43210, USA\\
$^{2}$NASA Sagan Fellow\\
$^{3}$Department of Astronomy and Astrophysics, University of Chicago, 5640 S Ellis Ave, Chicago, IL 60637, USA\\
fitzmaurice.11@osu.edu\\
}

\date{First submitted to MNRAS Dec 20, 2021}

\pubyear{2021}

\begin{document}

\label{firstpage}
\pagerange{\pageref{firstpage}--\pageref{lastpage}}
\maketitle

\begin{abstract}
Resonant locking of two planets is an expected outcome of convergent disc migration. The planets subsequently migrate together as a resonant pair. In the context of circumbinary planets, the disc is truncated internally by the binary. If there were only a single planet, then this inner disc edge would provide a natural parking location. However, for two planets migrating together in resonance there will be a tension between the inner planet stopping at the disc edge, and the outer planet continuing to be torqued inwards. In this paper we study this effect, showing that the outcome is a function of the planet-planet mass ratio. Smaller outer planets tend to be parked in a stable exterior $2:1$ or $3:2$ resonance with the inner planet, which remains near the disc edge. Equal or larger mass outer planets tend to push the inner planet past the disc edge and too close to the binary, causing it to be ejected or sometimes flipped to an exterior orbit. Our simulations show that this process may explain an observed dearth of small ($<3R_\oplus$) circumbinary planets, since  small planets are frequently ejected or left on long-period orbits, for which transit detection is less likely.  This may also be an effective mechanism for producing free-floating planets and interstellar interlopers like `Oumuamua.
   
\end{abstract}




\section{INTRODUCTION}\label{sec:introduction}

The discovery of thousands of exoplanets continues to challenge theories of planet formation. One of the most fundamental yet puzzling processes is disc-driven planet migration. For perhaps the majority of planets, it is not known if they migrated at all, with an alternative explanation being in situ formation. Hot-Jupiters are often exemplified as products of migration \citep{Dawson2018}, but there remains a debate of which type of migration, i.e. high-eccentricity \citep{Wu2007,fabrycky2007,Beage2012} or disc-driven \citep{lin1996,Kley2012}, in addition to lingering suggestions of in situ formation \citep{Batygin2015,Poon2021}. Resonant planetary systems are another class of planet for which migration is favoured \citep{Terquem2007,Ogihara2015,Inamdar2015,Raymond2018,Izidoro2021}. However, this presents a challenge because the process of convergent migration commonly produces resonances, yet in nature resonant systems are the exception, not the norm \citep{steffen2015}. This suggests that large-scale migration is either not a ubiquitous process \citep{Hansen2013,Chiang2013,Lee2016,MacDonald2020}, or there is a nearly equally ubiquitous process that can disrupt the majority of resonances \citep{rein2012PERIODRATIOS,Batygin2017}. In this paper we analyse a population of planets in which the history of migration is perhaps the least controversial: {\it circumbinary planets.} 

Twelve circumbinary systems (14 planets) have been found in transit, and the majority have a planet near the stability limit at $\sim 3a_{\rm bin}$ \citep{dvorak1984,holman1999,martin2014,li2016,quarles2018,Lam2018}. This is roughly where the protoplanetary disc would have been truncated by the tidal influence of the binary \citep{Artymowicz1994,miranda2015}. Repeated theoretical attempts have been made to produce circumbinary planets in situ.  The typical result has been that the highly turbulent environment near the inner disc edge is not conducive to planet formation \citep{Paardekooper2011,Lines2014,Meschiari2014,pierens2020,Pierens2021,Penzlin2021}.  Such challenges may be overcome with a very massive disc \citep{martin2013}, or formation that occurs in the late stages of the disc's lifetime \citep{childs2021,childs2022}, but the general consensus is that the known planets formed far away before migrating inwards \citep{Pierens2008,pierens2013,Kley2014,thun2018}. The steep surface gradient at the disc edge produces a repelling torque and causes the planet to ``park''. The effect of this natural stopping location on the observed distribution is clear -- it produces a ``pile-up'' of planets\footnote{Whilst the early analysis of \citet{martin2014} favoured the reality of a pile-up as opposed to an observational bias, the later work of \citet{li2016} suggested that more observations were needed. Since then three discoveries have been made (Kepler-1661, TOI-1338, TIC 172900988), all of which are near the stability limit. Irrespective of the propensity of planets to exist near the stability limit, it is widely believed that those found there did indeed migrate.}.

If circumbinary planets are indeed to be a golden sample of migrated planets, then it is imperative that we understand how migration will sculpt their architectures, and how we can best interpret the observed circumbinary distribution. One observed trend is a complete dearth of small ($<3R_\oplus$) circumbinary planets. This is in contrast to single stars, for which small planets (including super-Earths) are the most abundant planets known \citep{petigura2013,fulton2017}. This dearth could be caused by observational biases \citep{martin2018}. The binary causes transit-timing variations (TTVs) for circumbinary planets on the order of days to weeks \citep{armstrong2013} as compared to typically seconds and minutes for TTVs of single star planets \citep{agol2005,holman2005}. For single star planets we can stack the transits by phase-folding the light curve on an essentially fixed period and duration to enhance the signal and confirm a detection. The large TTVs of circumbinary planets mean that phase-folding on a fixed period and duration would instead wash away the signal. Consequently, the known planets were only detected because the individual transits are visible by eye. These hurdles mean that the dearth of small circumbinary planets could simply be a product of poor detection efficiency. Methods for alleviating these difficulties have been developed \citep{windemuth2019b,martin2021} but not yet widely applied. 

 The alternative explanation is that small planets either cannot form or survive around binaries. \citet{martinfitzmaurice2021} tested one theoretical mechanism for removing small  migrating circumbinary planets. The known planets are dispersed between potentially destabilising mean motion resonances with the binary, which the planets must have migrated through. Since migration speed is a roughly linear function of planet mass \citep{tanaka2002,lubow2010}, and slow migration makes resonant capture more likely \citep{Ketchum2011,Batygin2015,Mustill2016}, small planets may be preferentially imperiled. \citet{martinfitzmaurice2021} found that none of the known planets actually exist interior to a fully unstable resonance, defined to be one that is unstable even for circular planets. It was concluded that resonant ejection of slow-migrating small planets may occur in nature, but only if the disc is highly turbulent in order to pump up the planet's eccentricity. Overall it was determined that this effect would be unlikely to explain a complete dearth of small circumbinary planets.

In this paper we hypothesise and test a second mechanism for explaining a lack of small circumbinary planets: destabilising interactions in migrating, multi-planet systems. One multi-planet circumbinary system has been confirmed: the three-planet Kepler-47 \citep{orosz2012b,orosz2019}). The earlier work of \citet{kratter2014,smullen2016} provided a surprising result that multi-planet systems around binaries may be packed together almost as tightly as around a single stars. \citet{quarles2018} came to similar conclusions, and showed that in roughly half of the known systems an additional planet could be squeezed in between the known planet and the binary, despite the close proximity to the stability limit. However, these three studies only considered the stability of orbits already in place. \citet{sutherland2019} showed that in migrating two-planet systems the complex interplay of planet-planet and planet-binary resonances can lead to one or both planets being ejected. 

Our work builds upon the fundamental work of \citet{sutherland2019} in two primary ways. First, we include a prescription of a disc that is both truncated and turbulent, following the methodology of \citet{martinfitzmaurice2021}. This means that planet migration will not be constant and smooth, but rather will be stalled near the inner disc edge and have a degree of stochasticity. Second, we determine how interactions in this four-body system sculpt the size distribution of surviving planets, potentially explaining a dearth of small planets.

We open with a review of circumbinary planet migration, our N-body implementation of this effect, and mean-motion resonances (Sect.~\ref{sec:problem_setup}). We follow with demonstrations of how the planet mass ratio, planet starting positions, planet multiplicity, and disc parameters individually impact our simulations (Sect.~\ref{sec:demo_of_effects}). We then perform large population simulations to examine how the mass distribution of circumbinary planets is shaped by orbital interactions between companion planets (Sect.~\ref{sec:applications}). In Sect.~\ref{sec:discussion} we discuss the results of our simulations, including how analysis and detection of new circumbinary planets may be informed by our results, and a comparison of our results to certain single star systems before concluding in Sect.~\ref{sec:conclusion}.

\section{Problem Setup}\label{sec:problem_setup}

In this paper we implement the same methods as \citet{martinfitzmaurice2021} for modelling circumbinary planet migration within a truncated protoplanetary disc, with a key difference that this paper considers two-planet systems. In this section we briefly summarise the key concepts and equations, and refer the reader to \citet{martinfitzmaurice2021} for a more thorough summary.

\subsection{The picture of circumbinary planet formation and migration}\label{subsec:migration_picture}

The favoured paradigm for circumbinary planets is that they formed in the farther regions of the disc before migrating inwards and parking near a steep density gradient of a truncated protoplanetary disc \citep{Pierens2008,pierens2013,Kley2014,thun2018}. The alternative, in situ formation, is disfavoured due to a highly turbulent disc so close to the binary \citep{Paardekooper2012,Lines2014,Meschiari2014,pierens2020,Pierens2021}.

In our simple prescription of  migration there are two primary torques acting on the planet from the disc. Lindblad torques cause the planet to migrate inwards, and have a strength proportional to the disc surface density, $\Sigma$. Conversely, co-orbital torques cause outwards migration, with a strength proportional to the gradient of the disc surface density, $d\Sigma/dr$. The two torques balance when the planet is slightly interior to the peak of the disc density, where there is a steep density gradient due to the tidal truncation of the binary \citep{Artymowicz1994}. This causes the planet to ``park''. This mechanism has been demonstrated repeatedly in hydrodynamical simulations and is a foundation of both \citet{martinfitzmaurice2021} and this paper.

This picture is only applicable to Type-I migration, where the relatively low mass planet ($\lesssim0.5M_{\rm Jup}$) is embedded in the protoplanetary disc. Heavier planets carve a hole in the disc and migrate at a different, typically slower Type-II migration rate. For the known circumbinary planets near the stability limit it is believed that they all followed Type-I migration because if they cleared a hole in the disc then they would have removed all disc material within the co-orbital region, hence eliminating the outwards torque needed to park the planet. Such planets would migrate too close to the binary and ultimately be ejected \citep{Pierens2008}.

\begin{figure*}
    \centering
    \includegraphics[width=0.99\textwidth]{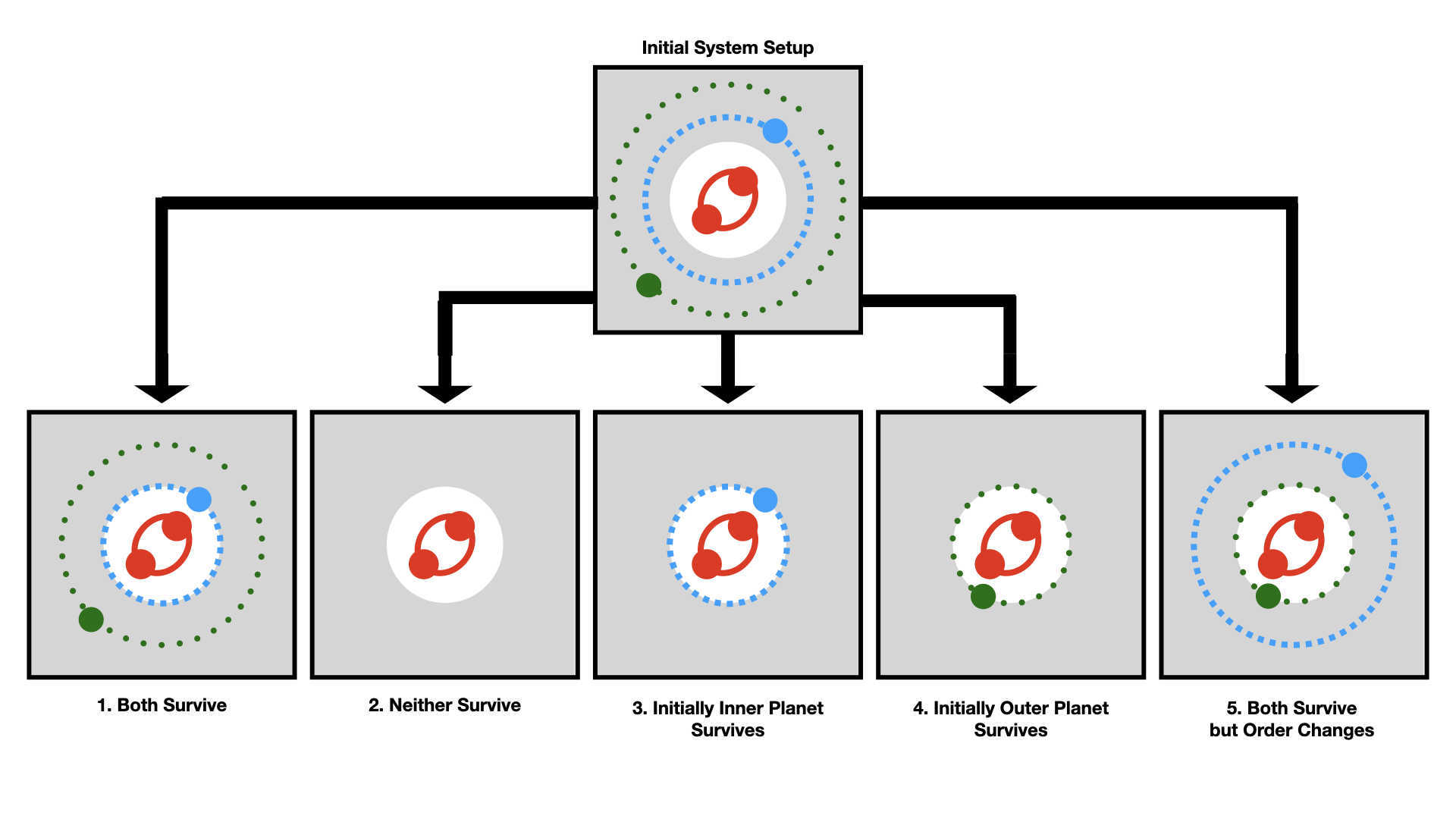}
    \caption{Five possible outcomes for final system architecture in our simulations. The figure is only meant to physically represent the order of the objects and the final position of the planets relative to the binary. The top panel shows the initial setup for all of our simulations: a binary (red, solid line) , an initially inner planet (blue, dashed line), and an initially outer planet (green, circle-dashed line). The grey backdrop signifies the disc, and the white circle surrounding the binary represents the region of the disc cleared out by the binary, which is closely approximated by the stability limit \citep{holman1999,quarles2018,Lam2018}. The bottom panels signify a different final outcome that may occur over varying timescales.}
    \label{fig:orbital_outcomes}
\end{figure*}

\subsection{N-body implementation of migration}\label{subsec:migration_nbody}

We use the \textsc{Rebound} N-body code \citep{rein2012REBOUND} with the WHFast integrator \citep{rein2015WHFAST} to model the circumbinary orbits. We opted to use WHFast over the IAS15 integrator because in \citet{martinfitzmaurice2021} we showed that with a sufficiently small time step $dt=0.001$ yr it could produce the same general results as the IAS15 integrator but with  significantly less
computation time. Migration is implemented using the \textsc{ReboundX} add-on package \citep{Tamayo2020} and its \textsc{modifyorbitsforces} where dissipative forces are created to mimic planet migration at a given rate $\tau_{\rm a}$. With a fixed $\tau_{\rm a}$ the planet orbit decays exponentially:

\begin{equation}
    \label{equ:a_p_function}
    a_{\rm p} = a_{\rm p,0}\exp\left(-\frac{t}{\tau_a}\right).
\end{equation}
The migration rate is related to the total torque from the disc $T_{\rm tot}$ by 
\begin{equation}
    \label{eq:tau_a}
    \tau_a = \frac{J_{\rm p}}{T_{\rm tot}},
\end{equation}
where $J_{\rm p} = m_{\rm p}\sqrt{a_{\rm p}G(m_{\rm A} + m_{\rm B} + m_{\rm p})}$ is the planet's angular momentum. The planet's eccentricity is also damped, at a rate $\tau_e$:
\begin{equation}
    \label{eq:tau_e}
    \tau_e = \frac{\tau_a}{K},
\end{equation}
where $K$ is a constant, with 10 being a typical value used by earlier studies \citep{lee2002,Kley2004}.

Our disc torques are calculated using prescriptions from \citet{tanaka2002,lubow2010}. The Lindblad torque is given by
\begin{equation}
    \label{eq:lindblad}
    T_{\rm L}(a_{\rm p}) = -\Sigma(a_{\rm p})\Omega_{\rm p}^2a_{\rm p}^4\left(\frac{m_{\rm p}}{m_{\rm AB}}\right)^2\left(\frac{a_{\rm p}}{H} \right)^2,
\end{equation}
where 
\begin{equation}
\label{eq:orbital_frequency}
    \Omega_{\rm p} = \sqrt{\frac{GM_{\rm AB}}{a_{\rm p}^3}}
\end{equation}
 is the Keplerian orbital frequency, $\Sigma(a_{\rm p})$ is the disc surface density (as a function of radius, approximated by $a_{\rm p}$) and $H$ is the disc height. For typical minimum mass solar nebula (MMSN) the disc height profile is $H=ha_{\rm p}$, for some constant aspect ratio $h$.
 
 The co-rotation torque is given by
 
 \begin{equation}
    \label{eq:coorbital}
    T_{\rm co}(a_{\rm p}) = s\Sigma(a_{\rm p})\Omega_{\rm p}^2a_{\rm p}w^3\left(\frac{\Delta\Sigma}{\Sigma} - \frac{\Delta B}{B}\right),
\end{equation}
where $B$ is the Oort constant, $B=\Omega_{\rm p}/4$ for a Keplerian disc, and $s$ is a factor we add to make ${\rm Max}(|T_{\rm L}|)={\rm Max}(|T_{\rm co}|)$, such that the co-orbital torque is guaranteed to balance the Lindblad torque. This balance parks the planet at the stability limit, one of our fundamental assumptions. Together, $T_{\rm tot}=T_{\rm L} + T_{\rm co}$ determines the migration rate in Eq.~\ref{eq:tau_a}.

These torques are calculated continuously throughout the simulation as the planet moves through the disc. The disc is modelled following hydrodynamical simulations such as \citet{pierens2013}:

\begin{equation}
    \label{eq:disc_profile}
    \Sigma(r) = f_{\rm gap}\Sigma_0r^{-3/2},
\end{equation}
where $\Sigma_0$ is a constant which determines the total disc mass, $r^{-3/2}$ is a typical power law and $f_{\rm gap}$ is used to model the truncation of the disc via

\begin{equation}
    \label{eq:f_gap}
    f_{\rm gap} = \left(1 + \exp\left[-\frac{r-R_{\rm gap}}{0.1R_{\rm gap}}\right] \right)^{-1}.
\end{equation}
The width of the disc's truncation region is determined by $R_{\rm gap}$, which will be typically near $\sim2.5-3.5$ and is the free variable which we use to force the known planets to park at their observed location.

All protoplanetary discs are expected to have a degree of turbulence. For a binary this will include traditional magnetorotational instability (MRI) turbulence, as well as stirring from the binary and eccentricity flucuations in the disc near the inner edge. This will add stochasticity to an otherwise smooth migration. We model this general turbulence using stochastic forcing, which was first developed by \citet{Rein2009,Rein2010PhD}, implemented in \citet{rein2012PERIODRATIOS,rein2012FOURTHREE} and similar to independent prescriptions used in \citet{Adams2008,Ketchum2011,Nesvorny2021}. 

Random forces are independently applied to the planet in the $x$ and $y$ directions, inducing a stochastic acceleration $a_{\rm SF}$ calculated by
\begin{equation}
    a_{\rm SF} = a_{\rm SF} * \exp \left(\frac{-dt}{\tau}\right)
\end{equation}
\begin{equation}
\label{eq:stoch_force}
    a_{\rm SF} = a_{\rm SF} + \left(\beta a_{\rm cent} \ X_i \ \sqrt{1 - \exp\left(\frac{-2\ dt}{\tau}\right)}  \right)
\end{equation}
where $dt$ is the change in time from the previous step, $\beta a_{\rm cent}$ is the amplitude of the stochastic acceleration, $\tau$ is the auto-correlation timescale and $X_i$ is a randomly generated scaling factor from a normal distribution with mean equal to 0 and standard deviation equal to 1  \citep{Kasdin1995,Rein2010PhD}. The amplitude of $a_{\rm SF}$ is defined relative to the amplitude of the acceleration due to the central binary $a_{\rm cent} = Gm_{\rm AB}/\left(3a_{\rm bin}\right)^2$, calculated at roughly the stability limit ($\approx 3a_{\rm bin}$). The scaling factor $\beta$ is a free parameter that we use to set the amplitude of the turbulence. \citet{rein2012PERIODRATIOS}\footnote{Although they label the parameter $\alpha$, but we avoid this notation since it may be confused with the classic \citet{shakura1973} $\alpha$ turbulence parameter.} tested a range of $\beta$ between $10^{-7}$ and $10^{-5}$ and \citet{Rein2009} estimate $\beta=5\times10^{-6}$. 

These random forces may change both $a_{\rm p}$ and $e_{\rm p}$. \citet{rein2012PERIODRATIOS,huhn2021} showed that stochastic forcing may break mean motion resonances, which in our study may protect planets from ejection. On the other hand, increased variation in $e_{\rm p}$ may precipitate instability. We follow \citet{rein2012PERIODRATIOS} in only applying stochastic forcing to one of the two planets. We apply it to the inner one, because we predict turbulence will be higher closer to the disc.

\subsection{Mean motion resonances}\label{subsec:resonances}

Mean-motion resonances can be the outcome of convergently migrating systems. \citet{martinfitzmaurice2021} analyzed the outcome of small circumbinary planets migrating through resonance with the binary (binary-planet resonances or BPRs). When considering migrating, multi-planet circumbinary systems, we must also consider resonances between the two planets (planet-planet resonances or PPRs). 

Two orbits will be in a mean motion resonance if their orbital periods are close to an integer commensurability, where ``close'' is defined by the resonant width, which is a function of the masses and eccentricities \citep{mardling2013}. A technical definition of resonance is that at least one of the resonant arguments $\phi$ librates, as defined by

\begin{equation}
\label{eq:resonant_argument}
    \phi = j_1\lambda_{\rm o} + j_2\lambda_{\rm i} + j_3\varpi_{\rm o} + j_4\varpi_{\rm i},
\end{equation}
where $\lambda$ is the mean longitude, $\varpi$ is the longitude of periapse and the $j$ coefficients are integers that follow the d'Alembert relation,
\begin{equation}
    \label{eq:dlamebert}
    \sum_{i=1}^{4}j_i=0
\end{equation}
(e.g. \citealt{murray1999}). When in a mean motion resonance $j_1=p+q$ and $j_2=-p$, and $j_3$ and $j_4$ come from Eq.~\ref{eq:dlamebert}.

Resonances play a key role in stability because  orbits in resonance undergo eccentricity variation. As the eccentricity of a body rises, it may move into resonant overlap zones, further intensifying eccentricity variation and resulting in ejection \citep{Mudryk2006}. In the four-body systems considered in this paper, there will be an interplay between PPRs and BPRs, as studied extensively by \citet{sutherland2019}. For example, in the Kepler-16 system the $6:1$ BPR is stable for $e_{\rm p}\lesssim 0.05$, as studied in detail in \citet{martinfitzmaurice2021}. A single planet can therefore safely navigate through this resonance as long as its eccentricity remains small. If the planet were instead in a PPR with an exterior companion, that PPR may increase the planet's eccentricity to the point where the $6:1$ BPR becomes unstable. 





\section{Demonstration of effects}\label{sec:demo_of_effects}
We use our N-body simulations to probe the effects of several parameters on the final architecture of initially multi-planet systems. Before analyzing the results of our larger simulations that encompass multiple parameters together, we use the following sections to demonstrate the individual effect of parameters such as planet mass ratio, planet starting position, disk density, and disk turbulence (stochastic forcing).

\subsection{Possible Outcomes}\label{subsec:outcomes}
Logically there are five possible outcomes for final system architecture and our simulations have produced them all. These five scenarios are displayed in Fig. \ref{fig:orbital_outcomes}. The figure is only meant to physically represent the order of the objects and the final position of the planets relative to the binary. 

The top panel shows the initial setup for all of our simulations, namely a binary (red, solid line), an initially inner planet (blue, dashed line), and an initially outer planet (green, circle-dashed line). The grey backdrop signifies the disc, and the white circle surrounding the binary represents the region of the disc cleared out by the binary \citep{Artymowicz1994}. This region fairly closely corresponds to the zone of instability \citep{holman1999,quarles2018,Lam2018}. The bottom panels signify a different final outcome that may occur over varying timescales.
\begin{enumerate}
    \item[\bf{1.}] {\bf Both survive:} both planets migrate inwards until the inner planet parks at disc edge and the outer planet remains on a stable, exterior orbit.
    \item[\bf{2.}] {\bf Neither survives:} there is an ejection of both planets, often almost simultaneously.
    \item[\bf{3.}] {\bf Initially inner planet survives:} the inner planet migrates and is parked at the inner edge of the disc, whilst the outer planet is ejected.
    \item[\bf{4.}] {\bf Initially outer planet survives:} the inner planet is ejected, leaving the initially outer planet to evolve as a single planet and park at the disc edge.
    \item[\bf{5.}] {\bf Both planets survive but swap order:} this occurs when the initially inner planet is flung to the outer regions of the disc but not fully ejected, causing a reversal in the two planets' distance from the binary.
\end{enumerate}

\subsection{Dependence on planet mass ratio}

\begin{figure}
    \centering
    \includegraphics[width=0.49\textwidth]{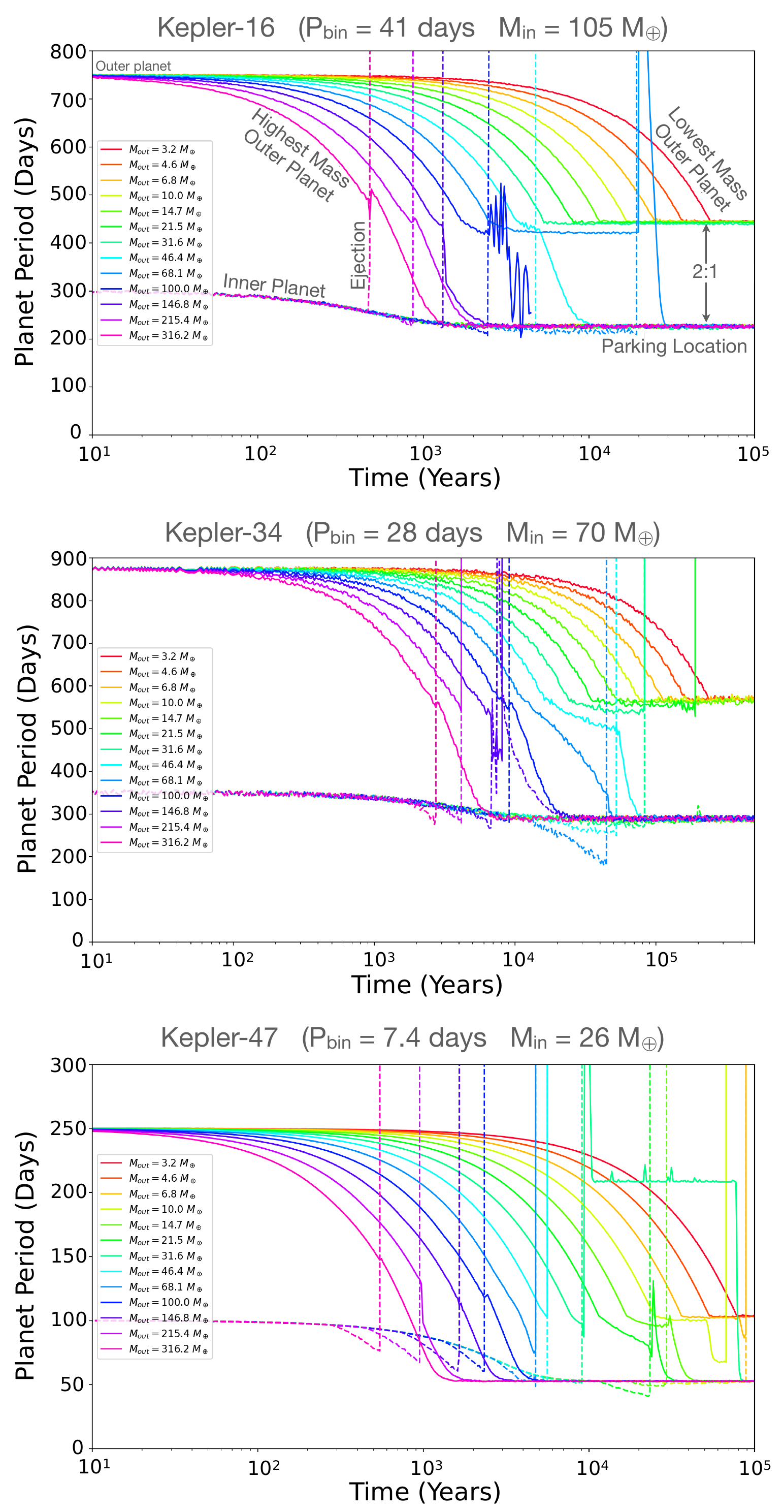}
    \caption{Simulations of three known circumbinary systems (Kepler-16, top; Kepler-34, middle; Kepler-47, bottom) with an added outer planet with mass between 3.2 and 316$M_\oplus$. Matching colours correspond to the same simulation, with the inner planet plotted with a dashed line. The outer period is initially 2.5 times the inner period. The inner planet mass is set to the known planet mass (for Kepler-47, the mass is not well constrained and this is roughly an upper limit). The disc parking location corresponds to the observed planet period (the innermost planet for Kepler-47 at 49 days). Significantly smaller outer planets tend to park at a stable $2:1$ exterior resonance. Significantly larger outer planets tend to eject the inner planet by forcing it too close to the binary. Planets of similar masses exhibit a range of behaviours, with one or both planets being ejected, or potentially trading places.}
    \label{fig:kepler_16_poster}
\end{figure}

\begin{figure}
    \centering
    \includegraphics[width=0.47\textwidth]{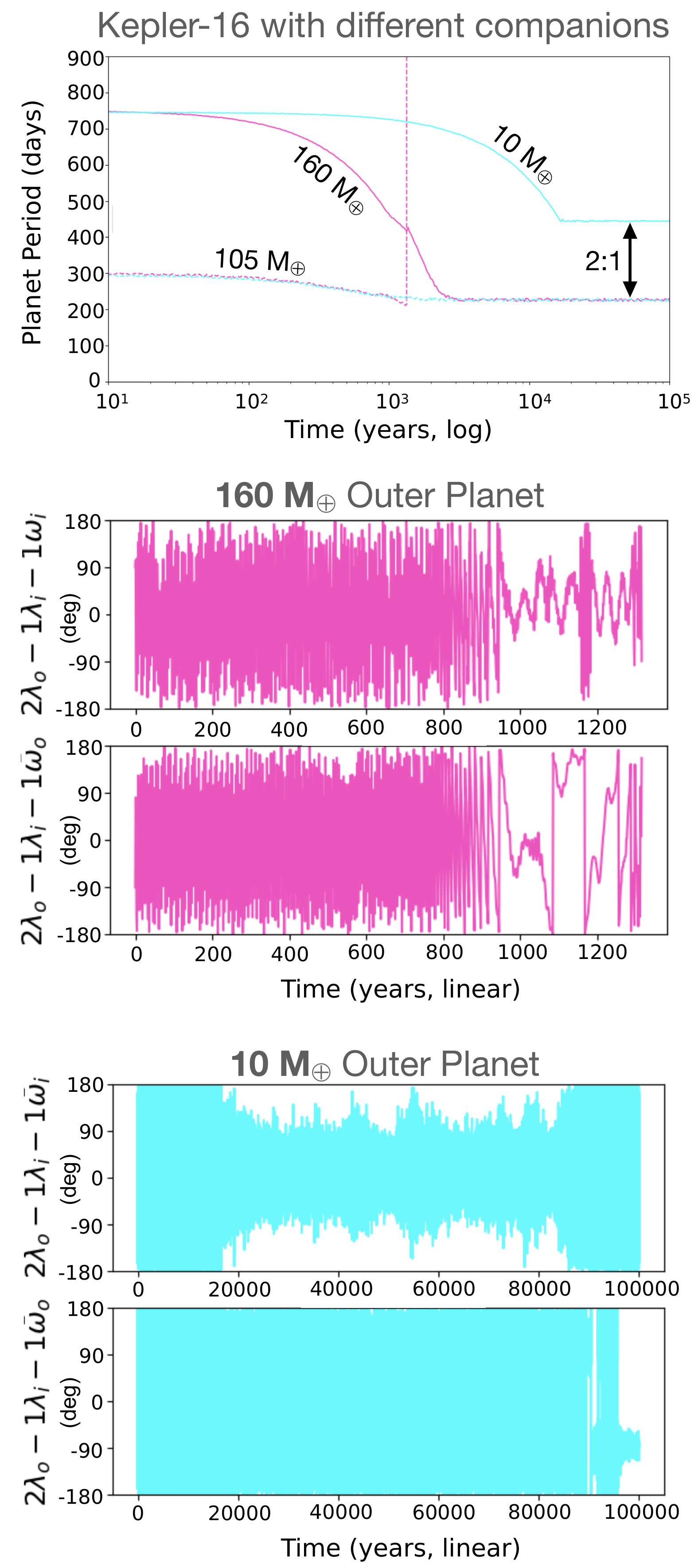}
    \caption{{\bf Top:} Migration paths of two separate two-planet systems around the Kepler-16 binary. The pink line represents a system with a more massive outer planet that leads to the ejection of the inner planet. The blue line represents a system with a less massive outer planet that  migrates in and lock into the $2:1$ resonance with the inner planet. Time is displayed on the x-axis is in log space. {\bf Middle, pink:} Resonant arguments of the 2:1 resonance for the more massive outer planet case. Shows variation of the argument but inability to lock into the 2:1 resonance. Time on the x-axis is displayed linearly and truncates around 1300 years when the inner planet is ejected. Note the linear time axis. {\bf Bottom, blue:} Resonant arguments of the $2:1$ resonance for the less massive outer planet case. Shows the libration of the second angle while the planets are locked into the $2:1$ resonance (top panel). After $\sim90,000$ years there is curiously a switch of libration between the two different resonant angles, but ultimately the planet is stable for the 100,000 year integration.}
    \label{fig:resonant_arguments}
\end{figure}

\begin{figure*}
    \centering
    \includegraphics[width=0.99\textwidth]{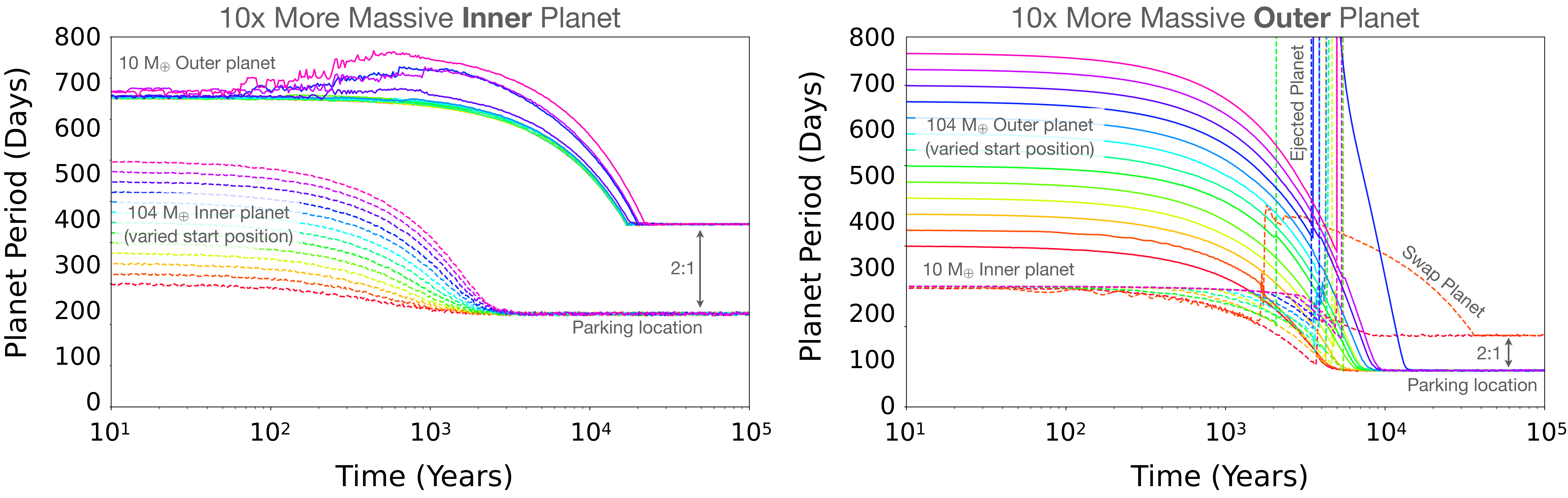}
    \caption{Simulations using the Kepler-16 binary and planet parameters to show the effect of starting position on the outcome in two planetary mass ratio cases. {\bf Left}: Varying the starting position of the Kepler-16b mass inner planet (dashed lines) with a 10 times less massive outer planet (solid lines). Corresponding colors represent the inner-outer pair. {\bf Right}: Varying the starting position of the Kepler-16b mass outer planet (solid lines) with a 10 times less massive inner planet (dashed lines). Corresponding colors represent the inner-outer pair. Starting position has little to no impact on the outcome of the system architecture in both regimes. }
    \label{fig:kepler_16_resonant_migration}
\end{figure*}

\begin{figure*}
    \centering
    \includegraphics[width=0.99\textwidth]{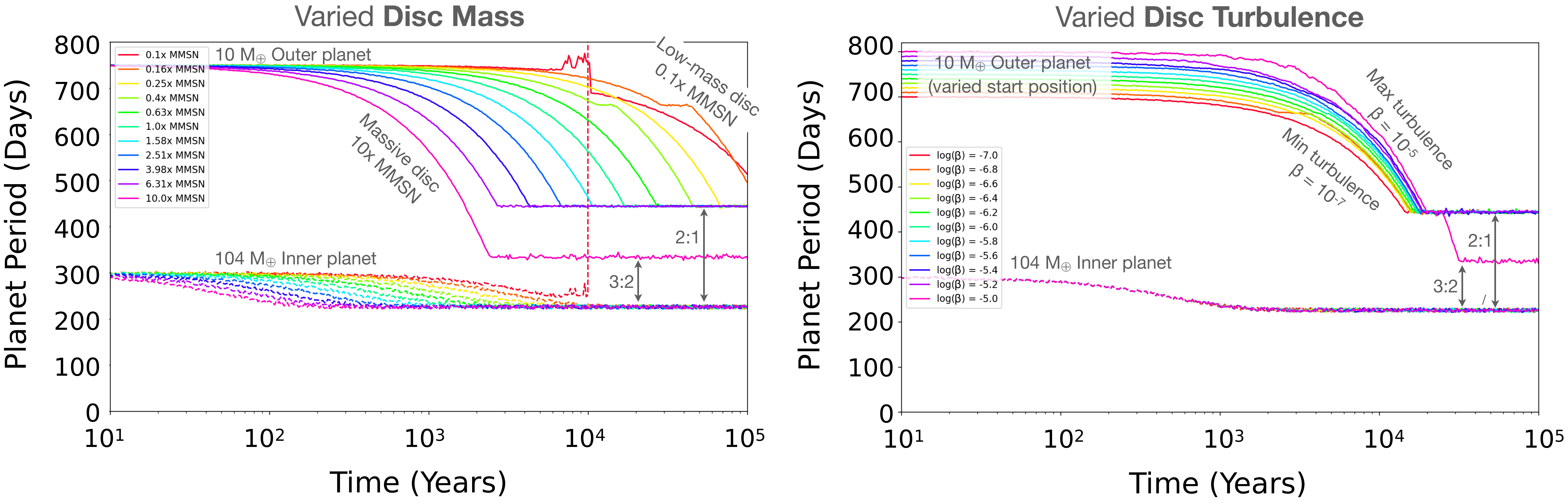}
    \caption{Simulations using the Kepler-16 binary and planetary parameters to show the effect of disc mass and turbulence on the system outcome. {\bf Left}: Varying disc mass for a Kepler-16b mass inner planet (dashed lines) and a 10 times less massive outer planet (solid lines). Corresponding line colors indicate planetary pair for a given disc mass from 0.1 to 10 times a Minimum Mass Solar Nebulae (MMSN). Disc mass alters migration speed, but is generally not disruptive on the outcome except in the least and most massive cases.  {\bf Right}: Varying disc turbulence through the strength of stochastic forcing for a Kepler-16b mass inner planet (dashed lines) and a 10 times less massive outer planet (solid lines). Corresponding line colors indicate planetary pair for a given strength of stochastic forcing for $\beta$ values of $10^{-7}$ to $10^{-5}$. The outer planet has a varied starting position to separate the outcome of each case visually. Stronger stochastic forcing can result in tighter orbital configurations in this planetary mass ratio case. }
    \label{fig:kepler_16_disc_variation}
\end{figure*}

We take three of the known circumbinary planet hosts: Kepler-16 ($M_{\rm A}=0.69M_\odot$, $M_{\rm B}=0.20M_\odot$, $P_{\rm bin}=41.08$ days, $e_{\rm bin}=0.16$), Kepler-34 ($M_{\rm A}=1.05M_\odot$, $M_{\rm B}=1.02M_\odot$, $P_{\rm bin}=27.80$ days, $e_{\rm bin}=0.52$) and Kepler-47 ($M_{\rm A}=1.04M_\odot$, $M_{\rm B}=0.36M_\odot$, $P_{\rm bin}=7.45$ days, $e_{\rm bin}=0.023$). These three binaries are chosen because they span a wide range of masses, mass ratios, periods and eccentricities. Around each binary we add two planets. The known planet is added with its observed planet mass and a starting orbital period 50 days longer than its observed orbit. A second, exterior planet is added at a period $P_{\rm out}=2.5P_{\rm in}$ and with 13 different tested planet masses, log-uniformly drawn between 3.2 and 316.2$M_\oplus$\footnote{A $316.2M_\oplus=1M_{\rm Jup}$ planet most likely would be in the Type-II migration regime, but we treat it as Type-I since the most important aspect of this test is the mass-ratio and not the absolute masses}. The parking radius is set to the observed planet's period and is what determines the disc profile through $R_{\rm gap}$ in Eq.~\ref{eq:f_gap}. We track the planet's eccentricity and consider it ejected when the value exceeds 1.

For this initial test we turn off turbulence (stochastic forcing) and use a $1\times$ minimum-mass solar nebula (MMSN) disc density  $\Sigma_0 = 1700$ g/cm$^2$ and scale height $h=0.04$. For Kepler-16 and -47 we run the simulation for 100,000 years. For Kepler-34 it is run for 500,000 years, since the longer periods of the planet mean that migration is slower.

The results are shown in Figure \ref{fig:kepler_16_poster}. Two effects are immediately obvious. First, the evolution of the system is dictated by the planet mass ratio. If the outer planet is heavier then it tends to act like a ``bulldozer'', meaning that it migrates inwards quickly without regard for the smaller inner planet (outcome 4 in Fig.~\ref{fig:orbital_outcomes}). This causes the inner planet to get pushed into the instability zone and be ejected, and then the initially outer planet migrates to the disc edge and parks. On the other hand, a significantly smaller outer planet gets stalled in the outer regions of the disc, leaving both planets stable (outcome 1 in Fig.~\ref{fig:orbital_outcomes}). If both planet masses are similar then it is seemingly stochastic what occurs, i.e. either planet or both of them may be ejected (outcomes 2, 3 and 4 in Fig.~\ref{fig:orbital_outcomes})

The second effect we see is that the $2:1$ PPR influences all of the outcomes. Consider a planet that has parked at the disc edge. Its migration has stopped because the disc torques are in equilibrium. When an exterior companion migrates into a $2:1$ period ratio, the two planets lock into a resonance. This locking happens regardless of the mass ratio. The effect of the mass ratio is to determine if the outer planet continues to migrate. A heavier outer planet continues to migrate because it feels strong a net inwards torque from the disc. The inner planet is locked into resonance, and is forced interior to the disc and too close to the binary, causing its ejection. On the other hand, if the outer planet is significantly lighter then when the $2:1$ PPR is reached the more massive, parked inner planet dictates the evolution. This causes the small outer planet to be parked at two times the period of the disc inner edge, which is why you see a ``pile-up'' of small outer planets along a horizontal line in Fig.~\ref{fig:kepler_16_poster}. Sometimes we see that a small outer planet locked into an exterior PPR gets ejected after a while. This is likely because planets locked in resonance have an increased eccentricity. For purely the PPR, this may remain stable, but when orbiting a binary this increased eccentricity may bring the outer planet into an unstable BPR, precipitating its ejection, as was found by \citet{sutherland2019}.

In Fig.~\ref{fig:resonant_arguments} we show the behaviour of the $2:1$ resonance in these two extreme cases by looking at the two resonant arguments. For the more massive outer planet (pink) $\phi_1=2\lambda_{\rm o} - \lambda_{\rm i} - \omega_{\rm i}$ becomes librating after $\sim1000$ years. This shows that the two planets have been locked in resonance. The inner planet migrates closer than its nominal parking location, and is ejected. For the less massive outer planet (blue) we interestingly see that at first $\phi_1=2\lambda_{\rm o} - \lambda_{\rm i} - \omega_{\rm i}$ librates, and then after $\sim 90,000$ there is a switch to $\phi_2=2\lambda_{\rm o} - \lambda_{\rm i} - \omega_{\rm o}$ librating, but this includes a period of about 3,000 years where neither resonant argument is librating. Regardless, the planet is ultimately parked at a period of $\sim460$ days.

\subsection{Dependence on starting position}

In the demonstrations in Figure \ref{fig:kepler_16_poster} we started our inner planet at a period just outside of its observed period, and the outer planet at a period 2.5 times that of the inner planet's initial period. Doing so neglects two potential scenarios. First, the planets may form at much greater distances, allowing them to migrate and lock into a resonance well before the inner planet reaches the stability limit. A second scenario is that the planets may start farther apart such that they could lock into higher-degree first-order resonances (e.g. $3:1$, $4:1$ and so on). We test the outcome of these scenarios in Figure \ref{fig:kepler_16_resonant_migration}. 

For a more massive {\it inner} planet (Fig.~\ref{fig:kepler_16_resonant_migration} left) the starting position does not change the final outcome. In each case the inner planet parks at the disc edge and the outer planet parks at the $2:1$ resonance. The only difference in the simulations is that when the inner planet initially starts farther out in the disc, it is closer to the outer planet and their interaction causes the outer planet to be pushed farther out.

For a more massive {\it outer} planet (Fig.~\ref{fig:kepler_16_resonant_migration} right) there is also qualitatively similar behaviour across most of the simulations, where the initially inner planet gets flung out from near the disc edge. The starting location of the more massive outer planet simply dictates when this interaction occurs. In most cases, this planet is ejected, but in two cases we see that the planet is captured in the disc. This small planet then migrates in and we have essentially the same setup as in (Fig.~\ref{fig:kepler_16_resonant_migration} left), as this small planet is parked at the $2:1$ resonance. This scenario where both planets are stable but swap positions is outcome 5 in Fig.~\ref{fig:orbital_outcomes}, but this appears to be a rare occurrence.

\subsection{Dependence on disc properties}

\begin{figure*}
    \centering
    \includegraphics[width=0.99\textwidth]{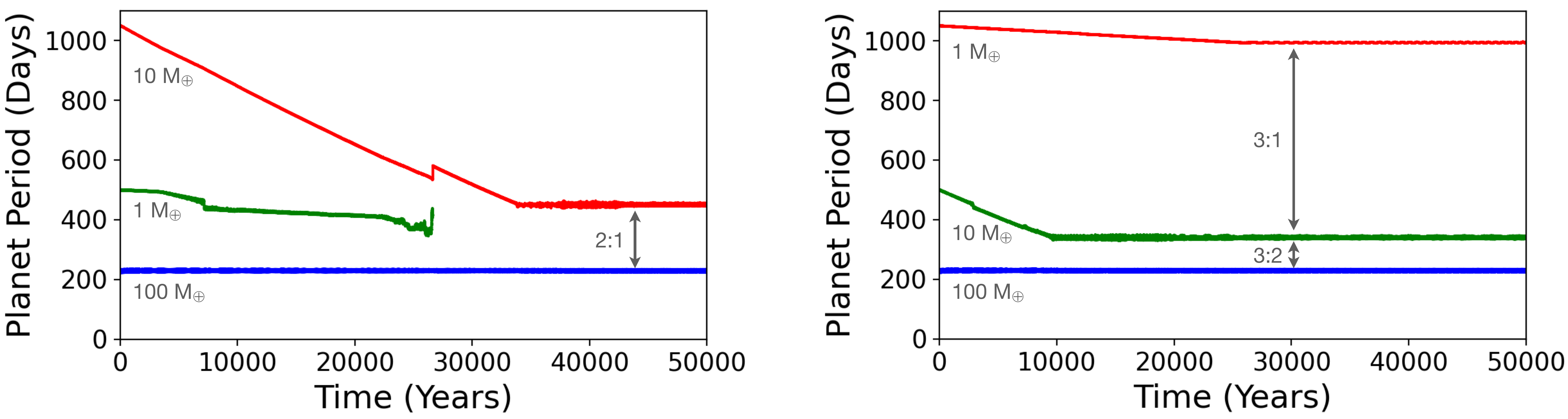}
    \caption{Kepler-16 with three planets. {\bf Left:} the middle planet is the smallest and is ultimately ejected. {\bf Right:} the outer planet is the smallest and a stable resonant chain is formed.}
    \label{fig:three_planets}
\end{figure*}

A more massive disc precipitates faster inwards migration, as is quantified by $\Sigma_0$ in Eq.~\ref{eq:lindblad}, which also means faster eccentricity dampening. This will affect resonant capture both between the two planets and between one planet and the binary. Heightened disc turbulence, quantified in our stochastic forcing model by $\beta$ in Eq.~\ref{eq:stoch_force}, may break apart resonances, and may also pump up planetary eccentricities to the point of instability.

In the left panel of Figure \ref{fig:kepler_16_disc_variation}, we show a variation of the disk mass from 0.1 to 10.0 times a MMSN for the Kepler-16 system with an added outer planet equal to 10$M_\oplus$. Stochastic forcing is turned off. For the least massive disc, we find that the inner planet migrates so slowly that it is locked into a $6:1$ BPR and is ejected, following the mechanism investigated in \citet{martinfitzmaurice2021}. We suspect that the outer planet at a period greater than 700 days did not have an effect here, but its exchange of angular momentum with the ejected inner planet does bring the outer $10M_\oplus$ body closer.

In the next step up in disc mass ($0.16\times$ MMSN), we see locking in the 3:1 resonance until the outer planet eventually breaks through and continues to migrate. For the 0.25 to $6.31\times$ MMSN cases we see migration through the 3:1 resonance and eventual locking into the 2:1 resonance resulting in stability through our simulations. For the most massive disk case, the outer planet migrates through the 2:1 resonance and ends up in a stable scenario locked in the 3:2 resonance with the inner planet. This variation in outcomes show that migration speed due to disk mass affects the resulting resonances that systems are able to lock into and preserve stable orbits. 

The right panel of Fig \ref{fig:kepler_16_disc_variation} shows variation of the parameter $\beta$ from Eq.~\ref{eq:stoch_force} which controls the level of disk turbulence. We vary $\beta$ from $10^{-7}$ to $10^{-5}$. This is again done on the Kepler-16 system with an outer planet equal to 10$M_\oplus$. We vary the starting position of the outer planet to visually separate the different cases. For smaller values of $\beta$ and thus lower turbulence, we find that the 2:1 resonance is still the catch-all resonance. We see a stalling in the 3:1 resonance for the $\beta = 10^{-6.8}$ but this is temporary and ultimately the planet moves to the $2:1$. For the highest level of turbulence, $\beta = 10^{-5}$, we see that the outer planet is able to break from the 2:1 resonance and migrate to a stable orbit at the tighter 3:2 resonance. This is the same result as found by \citet{rein2012PERIODRATIOS} for planets around single stars. Aside for this case, there is not a qualitative difference between the different levels of turbulence. The valuue of $\beta$  was increased to as high as  $10^{-4}$ in \citet{martinfitzmaurice2021} and  was found to cause destabilisation of all the planets in the simulations, leading to the conclusion that this value is too high. Turbulence likely plays a greater roll in the resonant ejection of slow-migrating single planets, as investigated in \citet{martinfitzmaurice2021}.

\subsection{Three planets}

In a three planet system the mass ratios of adjacent pairs will dictate the evolution. We again use the 41 day binary Kepler-16 as a test case, now with three added planets. The planets are started at $P_1=230$, $P_2=500$ and $P_3=1550$ days. This places the inner planet close to the parking position, and all adjacent pairs outside the $2:1$ resonance. The inner planet has a mass of $M_1=100M_\oplus$, similar to the real Kepler-16b. We test two configurations for the outer two planets: one with $M_2=1M_\oplus$ and $M_3=10M_\oplus$, and a second configuration reversed with $M_2=10M_\oplus$ and $M_3=1M_\oplus$. 

The results are shown in Fig.~\ref{fig:three_planets} for the two tested configurations. We see for $M_3 > M_2$ (left) the small sandwiched middle planet is caught into resonance with the faster-migrating outer planet, bringing it towards close contact with the inner planet. Ultimately the sandwich turns into a highly pressed panini, becoming too tight and ejecting the smallest, middle planet. For $M_3 < M_2$ (right) we are able to form a stable resonant chain, where each pairwise mass ratio is 0.1. Such size ordering is unlikely in nature, and requiring a 0.1 mass ratio for each pair would make it almost impossible to have four planets in a resonant chain.

\section{Applications}\label{sec:applications}

\subsection{Predicted outcome as a function of mass ratio}\label{subsec:probability}\label{subsec:probability}

\begin{figure*}
    \centering
    \includegraphics[width=0.99\textwidth]{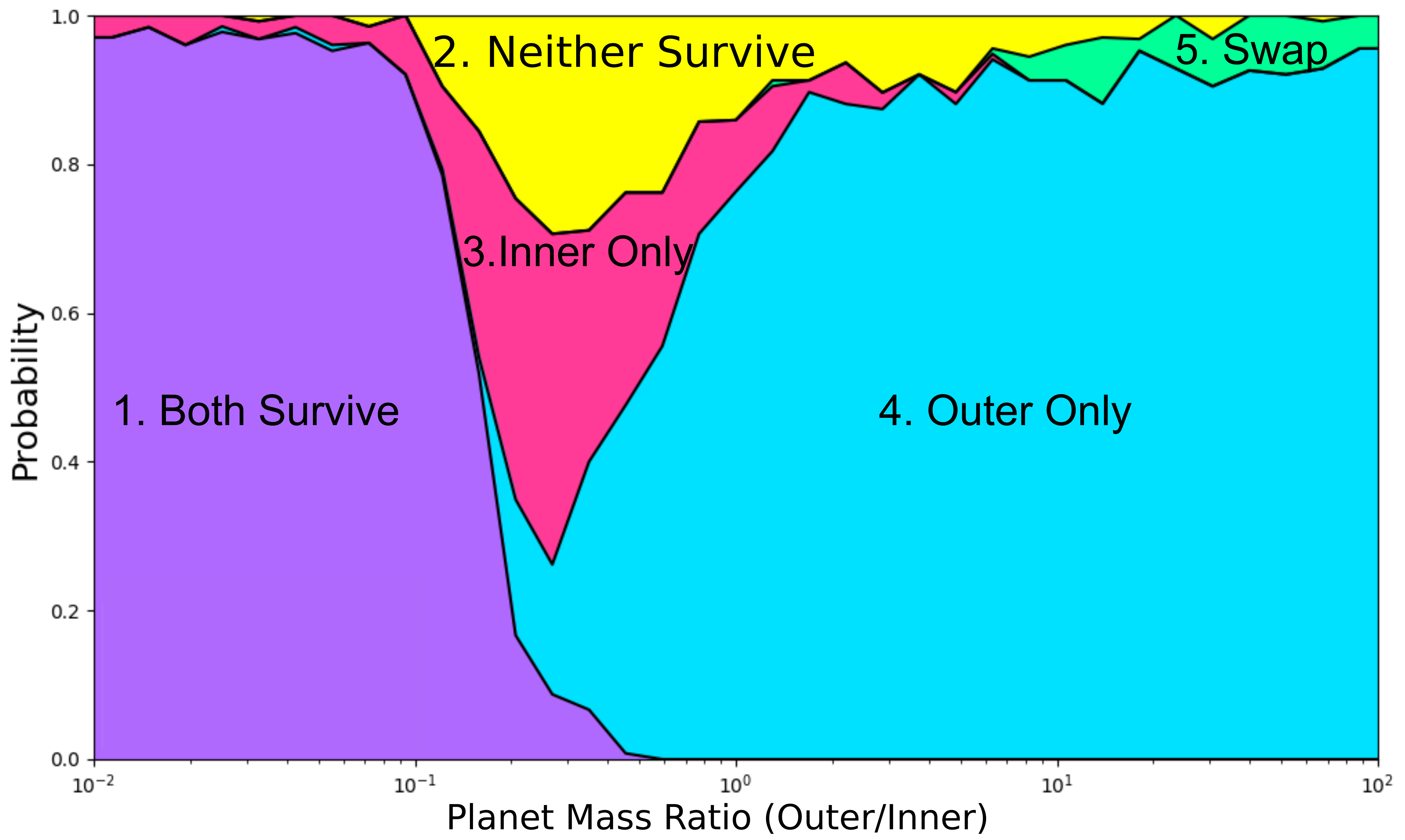}
    \caption{Outcomes of 100,000 year simulations of populations of two-planet circumbinary systems, where the systems are created to mimick the known transiting systems from Kepler and TESS but with a second, exterior planet added. The inner planet is started at 10\% from the orbital period of the known planet, which is set as the parking location. The outer planet is set at $P_{\rm out}=2.2P_{\rm in}$. The total planet mass is $150M_\oplus$ and 501 planet mass ratios are tested for each binary, on a log-uniform grid between $10^{-2}$ and $10^2$. The results from all of the different systems are combined on this plot to show regions of the planet mass ratio parameter space where the five possible different outcomes occur. The results are split up system by system in Fig.\ref{fig:known_trend}.}
    \label{fig:pop_trend}
\end{figure*}

From Sect.~\ref{sec:demo_of_effects} it is apparent that the planet mass ratio (outer/inner) is the greatest determinant in the outcome of the system. We ran a large suite of simulations to better quantify the probability of each of the five possible outcomes (Fig.~\ref{fig:orbital_outcomes}) as a function of planet mass ratio. The simulations were setup as follows. We take the known binary and stellar parameters\footnote{An up to date table can be found in \citet{martinfitzmaurice2021}.} for all 12 circumbinary planet hosts, with the exception of Kepler-1647 due to its very long planet period. The parking location is set to the known planet period (which determines the disc profile through $R_{\rm gap}$ in Eq.~\ref{eq:f_gap}). An inner planet is added 10\% away from this. An outer planet is added with $P_{\rm out} = 2.2P_{\rm in}$, i.e. just outside the important $2:1$ resonance. The planets are both added with circular orbits. We test a large array of 501 planet mass ratios $M_{\rm out}/M_{\rm in}$ on a log-uniform distribution between $10^{-2}-10^2$, where the total planet mass is kept constant at $150M_\oplus$. For the disc properties, $\Sigma_0=1700$ g/cm$^2$ ($1\times$MMSN) and we turn on stochastic forcing with $\beta=5\times10^{-6}$ \citep{Rein2009}.

Every simulation is run for 100,000 years, and we track ejections by seeing if the planet eccentricity becomes greater than 1. In Fig.~\ref{fig:pop_trend} we show the the probability parameter space for each of the five possible outcomes, as a function of the planet mass ratio.

For mass ratios less than 0.1, the vast majority of systems retain both planets. The configuration will be the more massive planet at the stability limit and the at least 10 times less massive planet parked at an external resonance. The chance of both planets surviving swiftly drops off at higher mass ratios. For intermediate mass rations between $\sim0.1$ and 2, i.e. for roughly similar mass planets, we see that either of the two planets can be ejected, or sometimes both. At higher mass ratios ($\gtrsim 2$) typically only the outer planet survives. This is because the less massive inner planet is pushed too close to the binary and ejected. A rare alternative is a ``swap'', where the less massive planet is pushed onto a wide orbit but not fully ejected. This process is seemingly stochastic and dependent on interactions between planets during the ejection of the inner planet.  In such cases the architecture effectively swaps, as we are in the very small mass ratio regime, for which both planets typically survive. 

\begin{figure}
    \centering
    \includegraphics[width=0.50\textwidth]{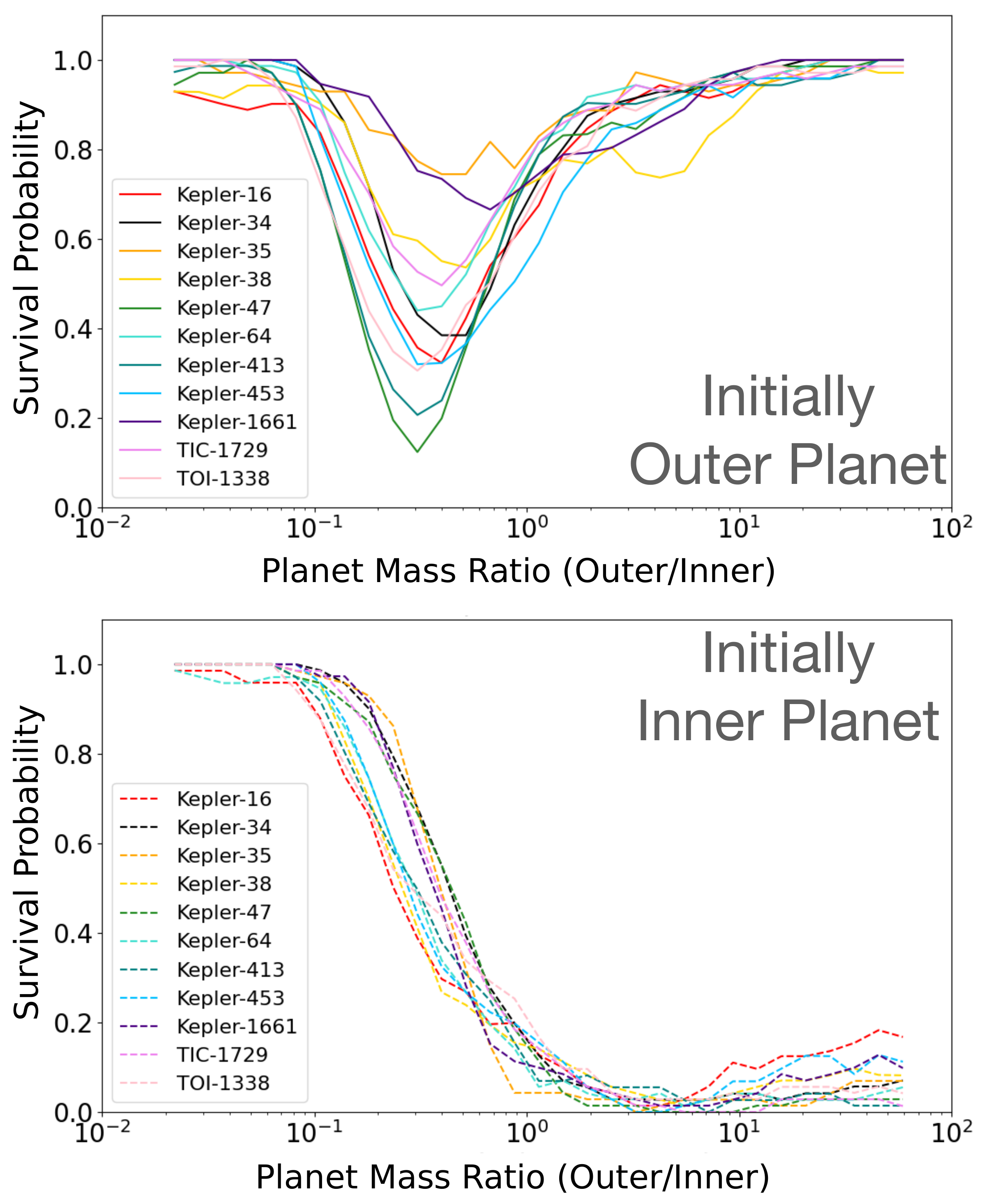}
    \caption{The results from the simulations in Fig. \ref{fig:pop_trend} split up into survival probabilities of the initially outer planet (top) and initially inner planet (bottom), also separated into all known transiting circumbinary hosts (except Kepler-1647 which hosts a 1100+ day planet).}
    \label{fig:known_trend}
\end{figure}

To better understand the transition region around intermediate mass ratios, in Fig.~\ref{fig:known_trend} we show the mean survival probability for the initially outer (top) and inner (bottom) planets, split up into the individually tested systems. The mass ratio is split into 36 bins (log-uniform spaced), and the survival probability is calculated with a rolling average over 5 bins.

For looking at both the inner and outer planets, we see that the width of the transition region where either of them may be ejected is roughly constant across all binary parameters, with bounds of roughly 0.1 to 2. Furthermore, for the inner planer the transition from full survival to nearly zero survival is also  roughly independent of the binary parameters, following a power law curve where the survival probability is inversely proportional to the mass ratio. For the outer planet whilst the width of the transition region seems fairly constant, the depth is not.

\subsection{Sculpting the size distribution}\label{subsec:sculpting}

The simulations of Sect.~\ref{subsec:probability} create a probabilistic model of the outcome of a migrating two-planet circumbinary system as a function of planet mass ratio. One can then generate any distribution of two-planet systems, apply the model of Fig.~\ref{fig:pop_trend}, and see what the predicted resulting distribution will be.

As a simple test, we construct a population of 1,000,000 two planets systems where each mass is independently drawn from a log-uniform distribution between $1M_\oplus$ and $1/2M_{\rm Jup}$. Whilst this (probably) is not a realistic representation of planets in nature, it demonstrates the effect on a flat distribution. For each of the 1,000,000 systems we calculate one of the five outcomes using the probabilities from Fig.~\ref{fig:pop_trend}. We then combine all surviving planets together, since the aim is to see the total impact on the circumbinary distribution. 

\begin{figure}
    \centering
    \includegraphics[width=0.50\textwidth]{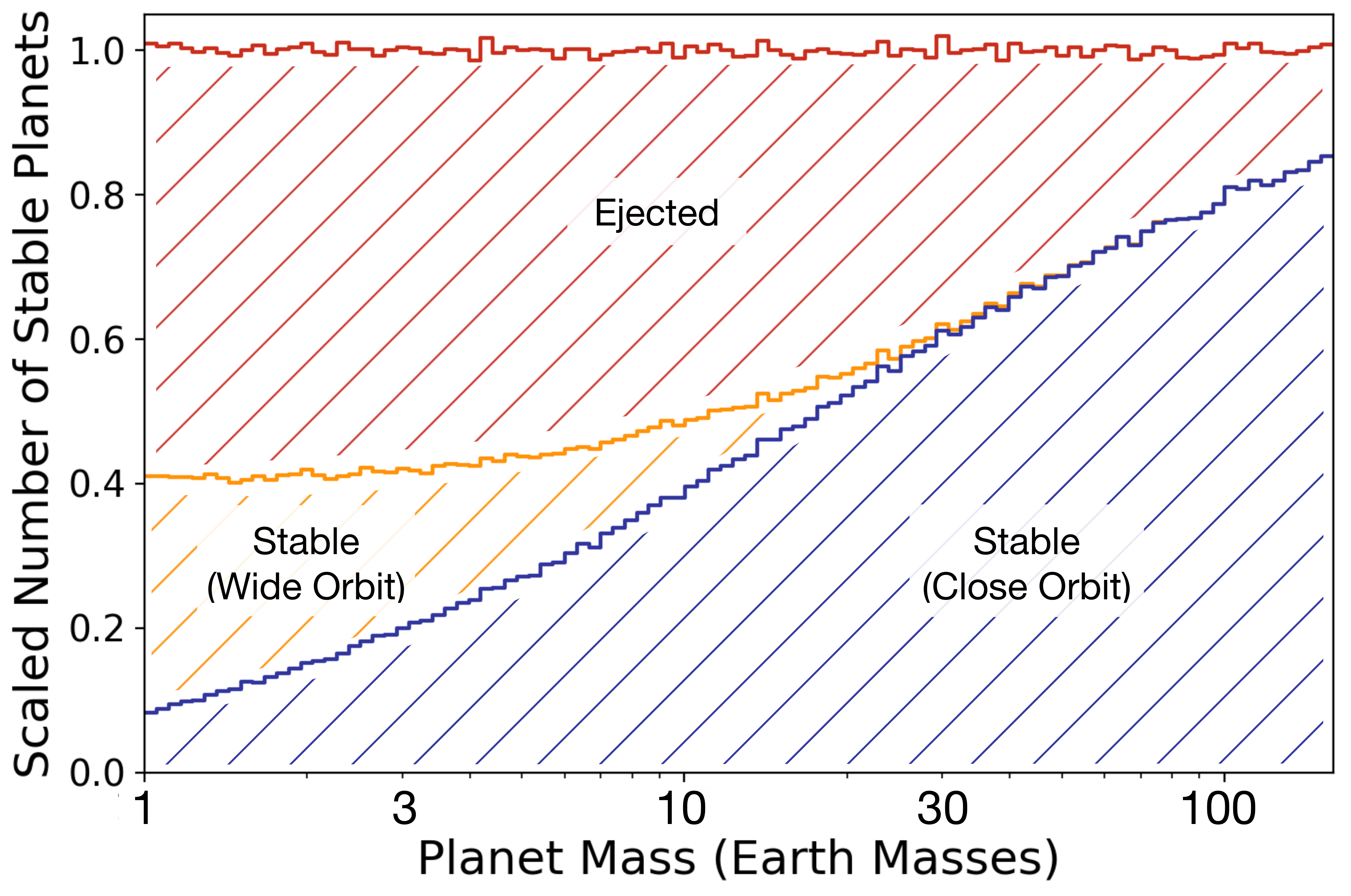}
    \caption{Input and output mass distribution of two planet circumbinary systems based on the probabilistic model of Fig.~\ref{fig:pop_trend}. The input distribution is log-uniform, and the number of stable planets is scaled so the input distribution is at 1. The red hatched region corresponds to all of the ejected planets. The blue hatched region is for planets that are stable and on orbits near the stability limit. The orange hatched region is for stable planets on wider orbits, locked in an external resonance with a larger interior planet. It is seen that multi-planet circumbinary systems sculpt the mass distribution to preferentially remove small planets.}
    \label{fig:sculpting}
\end{figure}

In Fig.~\ref{fig:sculpting} we show the sculpted mass distribution, where the number of stable planets is scaled such that the input flat distribution is at 1. The red  region refers to all of the planets that are ejected. The stable planets are split into two regions: blue for stable planets parked near the stability limit and orange of stable planets on wider orbits.

We see that for the most massive planets there is a $\approx20\%$ reduction in the planet occurence, but for the smallest planets it is a $\approx60\%$ reduction. The effect on smaller planets is even more profound if we only consider stable planets near the stability limt, for which there is a $\approx90\%$ reduction. This has observational consequences, because whilst the circumbinary planet transit probability is not as sharp a function of orbital distance as single stars \citep{martin2015,martin2017}, it will still be easier to find planets closer to the stability limit.

One thing to note is that the dynamics we test are a function of planet {\it mass}, whereas the transit method finds planets as a function of planet {\it radius}. In \citet{martinfitzmaurice2021} it is  noted that the \citet{Seager2007,Bashi2017} mass-radius relations give $10M_\oplus\sim 3R_\oplus$, and below this mass roughly half of the planets are ejected in multi-planet systems. Whilst there is significant scatter in the exoplanet mass-radius relationship, we nevertheless believe that many of the ``missing'' $<3R_\oplus$ planets may have been ejected due to interactions with a second planet.

\section{Discussion}\label{sec:discussion}

\subsection{Predicted circumbinary size distribution and architectures}

Our analysis shows that the evolution of a migrating two-planet circumbinary system can be split into three categories based on the planet-planet mass ratio. For $M_{\rm out}/M_{\rm in} > 0.3$, only one planet typically survives, and it is the outer, more massive one. A rare exception to this is when the small, initially inner planet is not quite ejected, but rather swaps positions and becomes a stable, smaller exterior planet. For $M_{\rm out}/M_{\rm in} < 0.05$ both planets survive, with the less massive, outer planet parked at a resonance, typically the $2:1$. In between these two mass ratios the outcome is seemingly stochastic, with one or both planets being ejected. 

For the circumbinary size distribution, we predict a legitimate dearth of small planets due to destabilising interactions with a companion planet. This is arguably a more efficient mechanism to ejection of slowly migrating single planets in \citet{martinfitzmaurice2021}. The \citet{martinfitzmaurice2021} mechanism does not require a second planet, but it does require a highly turbulent disc. Whilst the mechanism put forward in this current paper does require a second planet, it works over a large range of planet-planet mass ratios, including for similar-mass planets. Furthermore, the underlying exoplanet multiplicity around single stars is $64\%$ \citep{sandford2019} and at least one known multiple circumbinary system is known (Kepler-47 with three planets, \citealt{orosz2012b,orosz2019}). 

For the architectures of circumbinary planets we predict that planets found near the stability limit will be unlikely to have a more massive exterior companion interior to the $2:1$ (or possibly $3:2$) PPR. For smaller exterior companions we predict that they will be either locked into one of these first order resonances, or on a wider orbit. 

One big topic of study around single stars is the ``peas in a pod'' phenomena, where multi-planet systems are often found with regular sizes and spacing \citep{weiss2018,wei2020,jiang2020,millholland2021}. In our simulations similar mass systems almost always result in at least one planet being ejected. That would suggest that the size regularity would be rare around binaries, at least when the planets are close to the stability limit. For three or more planets, we show that to form a stable resonant chain near the stability limit each consecutive planet mass ratio needs to be $M_{\rm out}/M_{\rm in} \lesssim 0.1$. Such a configuration is likely rare in nature, and hence we suggest that the orbital spacing regularity of peas in a pod may also be rare in circumbinary systems.


\subsection{The only known multi-planet circumbinary system -- Kepler-47}\label{subsec:kepler47}
Kepler-47 ($M_{\rm A}=1.04M_\odot$, $M_{\rm B}=0.36M_\odot$, $P_{\rm bin}=7.45$ days, $e_{\rm bin}=0.023$) is the only binary system known to host multiple planets. The planets in order from inner to outer are Kepler-47b ($M_{\rm b} \leq 25.77 M_{\oplus}$,  $P_{\rm b} = 49.53$ days), Kepler-47d ($M_{\rm d} = 19.02 M_{\oplus}$, $P_{\rm d} = 187.35$ days), and Kepler-47c ($M_{\rm b} = 3.17 M_{\oplus}$, $P_{\rm b} = 303.14$ days) \citep{orosz2012b,orosz2019}. The inner planet's mass only has an upper limit, and its radius at $3R_\oplus$ is the smallest of the three.

The inner two planets have a period ratio of 3.78, which fits our prediction that there should not be planets interior to a $2:1$ or $3:2$  resonance. The simplest explanation for why the middle planet is so far away is that the disc dissipated before it had time to migrate in closer. for the outer pair the period ratio is 1.62. This interior to the $2:1$ period ratio which we typically see planets lock into in our simulations. Furthermore, one would have expected the $82\times$ more massive middle planet to migrate in much quicker than the outer planet, increasing their separation. Overall, this predicts that the outer two planets likely formed even closer together. A more exotic alternative is that the small outer planet was brought interior to the $2:1$ PPR by being pushed in by a fourth, outer planet, although the dynamical signature of such a planet would likely be visible in the transit timing variations.

\subsection{Comparison with past circumbinary work}\label{subsec:comparison_past_work}

The earlier work of \citet{kratter2014,smullen2016,quarles2018} showed that stability of multi-planet systems around binaries was somewhat similar to that around single stars, as long as you are exterior to the inner instability zone. However, this was based on a static (i.e. non-migrating) orbits. \citet{quarles2018} showed that  Kepler-34, -47, -413, -453 and -1647 could host an interior planet in between the known planet and the stability limit\footnote{According to \citet{quarles2018} Kepler-16, -35, -38 and -64 could {\it not} host an interior planet.}. Whilst this would be true if the additional planet were placed there, our work and  that of \citet{sutherland2019} shows that to migrate planets to such a tightly packed system near the stability limit would be very difficult. 

\citet{gong2017} studied scattering of circumbinary planets in the late stages of formation (i.e. when the gas has disappeared, so after any migration). Similar to our work, they showed that small planets are typically the ones ejected.

Our N-body prescription of circumbinary planet migration is constructed to replicate the general results of  more complex hydrodynamical simulations such as \citet{Pierens2008,pierens2013,Kley2014,thun2018,Penzlin2021}. All of these studies focused on single planet circumbinary systems. To our knowledge, \citet{penzlin2019} has been the only hydrodynamical simulation applied to multi-planet circumbinary systems. They studied Kepler-47 and -413 and showed that two equal mass small planets can actually both migrate to the inner edge of the disc and be captured into a $1:1$ PPR. The two planets end up on the same orbital period in a horseshoe configuration. In our simulations we found that such a situation occurred very rarely, but it may be a situation that is more likely if the hydrodynamical subtleties of discs are studied. \citet{penzlin2019} also note that their two studied binaries are the most circular known ones ($e_{\rm bin}<0.04$). When they applied their work to the more eccentric Kepler-34 and Kepler-35, one of the planets was ejected, similar to what we see.

\subsection{Comparable effects around single stars} \label{subsec:comparison_single_stars}
Whilst binaries carve out a hole in their  disc via gravitational interactions, single stars may have a similar effect on their discs via magnetic interactions.  \citet{lee2017} describes the truncation at corotation of discs around FGKM stars based on magnetospheric interactions and the stellar rotation period. \citet{Ataiee2021} conducted a similar study to ours of migrating resonant planet pairs in the presence of a truncated inner disc boundary. For this type of boundary, \citet{Ataiee2021} finds that resonant chains of 2 or more planets, when mass increases with increasing semi-major axis, can push the interior most planet past the density buildup at the disc truncation. In the circumbinary case, we know this forced migration past the disc truncation leads to the ejection of this planet. In the single star case, such migration may not be as fatal and could lead to the creation of short and ultra-short period planets . Alternatively, the planet could crash into the star. \citet{Ataiee2021} also study the same effect but near ``dead zones'' in disc, which in their model correspond to an even steeper density gradient. In line with our model, having a steeper gradient strengthens the outwards co-orbital torque. They conclude that resonant pushing of planets past a dead zone is less likely to occur than for an inner boundary. 

\citet{HuangTrappist2021} conducted a similar study of the resonant chain migration of the TRAPPIST-1 system. They saw two effects. First, the inner planet was likely pushed interior to the inner disc edge, through resonant interactions. Second, some of the outer planets were likely ``piled up'' in this resonant chain, in a way similar to what we see in our simulations.

\subsection{Free Floating Planets}\label{subsec:free_floating_planets}

Free-floating planets are ``rogue'' planets without a host star. Such planets were first discovered roughly a decade ago (review in \citealt{gaudi2012}).  \citet{sumi2011} produced a surprising discovery that there were roughly two Jupiter-mass planets for every star in the Galaxy, although later work by \citet{mroz2017} challenged this and lowered the rate to 0.25 rogue planets per star. With the upcoming Nancy Grace Roman Space Telescope (Roman) and its Galactic Exoplanet Survey, we will be sensitive to potentially hundreds of free floating planets, with masses down to that of Mars \citep{johnson2020}.

A leading means of producing these planets is through scattering. \citet{veras2012} said that this mechanism was not sufficiently efficient, but this was based on the potentially inflated predictions of \citet{sumi2011}. \citet{Sutherland2016,smullen2016} both showed that the fate of unstable planets around binaries was typically ($\sim 70-80\%$) ejections, as opposed to collisions with either star or collisions with another planet. In \citet{smullen2016} they make a comparison with planet-planet scattering around single stars, showing that in the single star case ejections occurred less than roughly half of the time. Those studies though had static (i.e. non-migrating) populations of planets. Here we show that two migrating planets in a truncated circumbinary disc frequently result in one or both of the planets becoming unstable. Between this discovery and earlier work in \citet{martinfitzmaurice2021} on the instability of slowly-migrating single circumbinary planets, we have two new mechanisms for creating a population of rogue planets. 

 The escape velocities from binary and single star systems are expected to be different \citep{levine2021}. As a future study it would be interesting to quantify the expected differences on populations of free-floating planets, which may be testable by Roman. care will be needed to not confuse these planets with ejected planets from other fast stellar sources. For example, halo stars typically have higher velocities (e.g. \citealt{li2021}) and they are typically older, meaning that dynamical instability had longer to arise. Presently, though, our observational constraints on halo planets are in their infancy \citep{kolecki2021,boley2021}.


\subsection{Interstellar interlopers}\label{subsec:discussion_interlopers}

In a similar vein to free floating planets, the recent discoveries of the interstellar interlopers `Oumuamua and Borisov have prompted debates over their origin.  \citet{jackson2018,cuk2018} proposed that `Oumuamua may be the result of ejection from a binary star system because the ratio of ejections to accretion is much higher if the secondary object is a star rather than a planet. In \citet{jackson2018} they migrate in planetesimals with an assumption that an amount ($\sim 10\%$) will not pile up at the steep density gradient, and rather pass beyond the critical stability limit and be ejected from the system.

\begin{figure}
    \centering
    \includegraphics[width=0.49\textwidth]{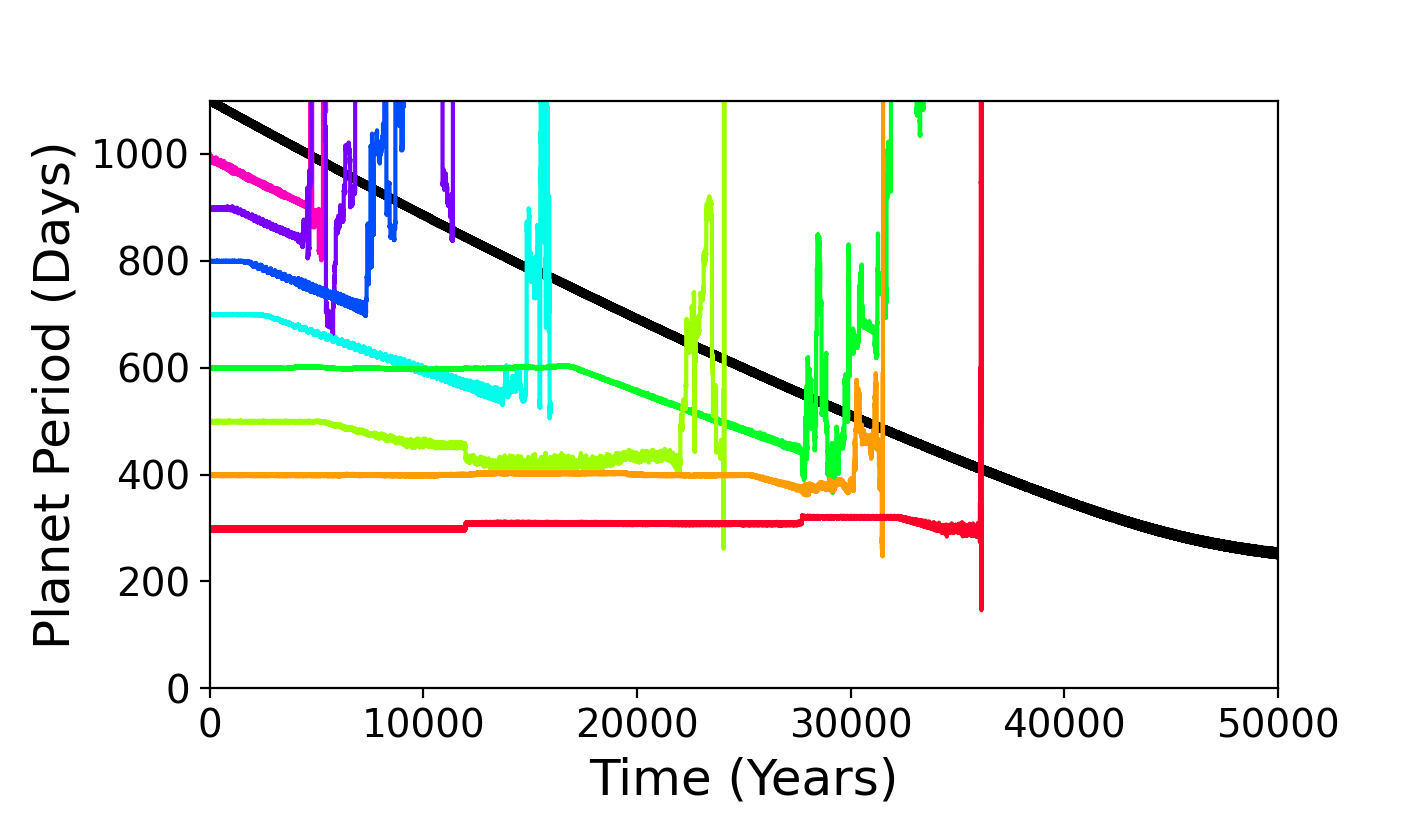}
    \caption{Inwards migration of a $10M_\oplus$ planet (black) through seven evenly spaced $0.01M_\oplus$ bodies (colours), orbiting the Kepler-16 binary. The planetesimals would themselves migrate very slowly, but they are swept up into resonances by the migrating planet, ultimately coming close to the binary and other bodies and being ejected. This may be a mechanism for flooding the galaxy with small bodies like `Oumuamua.}
    \label{fig:planetesimals}
\end{figure}

In our paper we provide a natural means of pushing small planetesimals beyond the critical limit through resonant interactions with a larger migrating planet. We demonstrate this in Fig.~\ref{fig:planetesimals}. We see an exterior planet  migrating through the disc towards the disc edge. It sweeps up interior smaller bodies into resonances, ultimately resulting in their ejection.  Given that the abundance of gas giants is roughly similar around single and binary stars \citep{armstrong2014,martin2014,martin2019}, and it is believed they migrated to their current location near the stability limit, this resonant sweeping may be an efficient production line for interstellar interlopers. 

 \citet{levine2021} note that the non-ballistic trajectory of `Oumuamua calls into question a circumbinary origin. An alternative may be ejection from an evolving single star system \citep{katz2018}. As our sample of interstellar interlopers grows, so will our knowledge of their origin. 

\subsection{Observational tests} \label{subsec:observational_tests}

\subsubsection{Transits}

\begin{table}
\begin{center}
\label{tab:detection_limits}
\begin{tabular}{ccccc}
\hline
Name        & Orbital  & True  &    Detection  & Detection \\
        &  Period & Radius  &     Limit &  Limit\\
            & $P_{\rm p}$ (days) &  ($R_{\oplus}$) &   at $P_{\rm p}$ ($R_{\oplus}$) &at $2\times P_{\rm p}$ ($R_{\oplus}$) \\
\hline
Kepler-16   & 228.78 & 8.27         & 0.79            & 0.94                       \\
Kepler-34   & 288.82 & 8.38         & 3.81            & 4.36                       \\
Kepler-35   & 131.46 & 7.99         & 4.39            & 5.22                       \\
Kepler-38   & 105.60 & 4.20         & 3.88            & 4.61                       \\
Kepler-47b  & 49.53 & 3.05         & 2.20            & 2.61                       \\
Kepler-64   & 138.32 & 6.10         & 4.29            & 5.05                       \\
Kepler-413  & 66.26 & 4.35         & 3.56            & 4.23                       \\
Kepler-453  & 240.50 & 6.30         & 1.57            & 1.87                       \\
Kepler-1661 & 175.05 & 3.87         & 2.49            & 2.96            \\
\hline
\end{tabular}
\caption{Detection limits for Kepler-circumbinary systems at the known planetary period and at the exterior $2:1$ PPR based on the \textsc{Stanley} algorithm (\citealt{martin2021}, Fig. 14 specifically).}
\label{tab:detection_limits}
\end{center}
\end{table}

The STANLEY algorithm  presented in \citealt{martin2021} uses brute-force grid search and N-body integration to automatically stack shallow circumbinary planet transits and reveal  planets much smaller than the known by-eye detections. In Table~\ref{tab:detection_limits} are the radius and orbital period for the known Kepler circumbinary planets\footnote{With the exception of Kepler-1647, Kepler-47c and Kepler-47d, which are detectable by \textsc{Stanley} but detection limits were not calculated for.}, the \textsc{Stanley} detection limit at the known period and the \textsc{Stanley} detection limit at the $2:1$ external PPR. In all cases \textsc{Stanley} could detect a similar or smaller radius planet at the $2:1$ resonance. In 4/9 cases we could detect a $< 3R_\oplus$ planet, i.e. the type of planet not found so far. We note that while a planet could be detected in all of these cases, only Kepler-47 is known to host exterior planets. A lack of outer planet detections in these systems could signify that multi-planet circumbinary systems typically have smaller planets on the interior, and they are subsequently ejected when the gas giant migrates inwards and are rarely re-captured in exterior orbits. A significant increase in detections of circumbinary systems will determine the significance of the orbital interactions we have explored. 


\subsubsection{Radial velocities}

The BEBOP  survey  has been searching for circumbinary planets using radial velocities since 2014. The advantage of radial velocities over transits in the context of this work is that planets are found as a function of mass, not size, and it is their mass that determines their migration rate. The first survey results were presented in \citealt{martin2019}, based on 4 years of observation of 48 binaries with the CORALIE spectrograph. Whilst the survey was limited as a function of mass to typically Jupiter-sized objects, the  radial velocity method is not as sharply dependent on orbital period as transits, and hence there was significant parameter space for multi-planet giant systems to be found. The survey has since been upgraded to the HARPS, SOPHIE and ESPRESSO spectrographs. \citet{triaud2021} recovers the radial velocity signal of Kepler-16b, the first such time a circumbinary planet has been observed with this technique, validating this method. Preliminary results  in \citet{standing2021} show a sensitivity almost down to Neptune mass in orbits near the binary. Saturn mass planets could be found at  longer period orbits out to $\sim3-6$ years, well away from the stability limit.  As the BEBOP observing baseline reaches a decade and beyond, we will ultimately be able to interpret the results in the context of multi-planet circumbinary systems.

\subsubsection{Gravitational microlensing}

Microlensing  is arguably the most sensitive to small planets, and its observational sweet spot is at typically a few AU, making it complementary to transits and radial velocities \citep{gaudi2012}. With the upcoming Nancy Grace Roman Telescope, there is tremendous growth expected for this field. Microlensing has been applied to circumbinary planets \citep{luhn2016,george2021}, albeit with the complexity of two signals to interpret. One planet has been found so far: OGLE-2007-BLG-349L(AB)c \citep{bennett2016}. It awaits to be seen if a multi-planet circumbinary system could be detected with microlensing, but even with single-planet systems this method may offer a window into the mass distribution of planets farther out from the binary.



\section{Conclusion}\label{sec:conclusion}

In this paper we investigate the stability and orbital evolution of two planet circumbinary systems. In particular, we examine the curious interplay of mean motion resonances (both binary-planet and planet-planet) and the truncation of the protoplanetary disc. We use an N-body prescription of migration in a truncated and turbulent circumbinary disc, based on  previous work in \citet{martinfitzmaurice2021}. Our simulations show that planets have a tendency to lock into mean motion resonances, and subsequently migrate together as a locked pair. When the innermost planet reaches the stability limit the outcome is largely dictated by the mass ratio. If the outer planet is significantly less massive ($M_{\rm out}/M_{\rm in}\lesssim0.2$) then typically the inner planet parks at the disc edge and the outer planet parks at a  stable external resonance (typically $2:1$ or $3:2$). If the outer planet is significantly more massive ($M_{\rm out}/M_{\rm in}\gtrsim2$) then it behaves like a bulldozer, forcing the inner planet into the instability zone and causing it to be ejected. The more massive, remaining planet then migrates into the disc edge and parks. In this scenario occasionally the less massive planet is not ejected, but rather the two planets swap their positions. For intermediate mass ratios, where the two planets are closer in mass, we see that one or both planets get ejected, and the ultimate result is effectively stochastic. 

With respect to small planets, there are two robust trends  seen in our work. First, the planets that are ejected are preferentially small. Second, when small planets remain on stable orbits they typically do so on long-period orbits, locked in resonance with a more masive, interior planet. Both effects work to explain a dearth of small ($<3R_\oplus$) circumbinary planets \citep{martin2018}, as these planets may either not exist or exist on long-period orbits, harder to find with transits.  This mechanism may also provide efficient production of free-floating planets and interstellar interlopers like `Oumuamua. We also find that stable multi-planet systems with two similar mass planets are rare, meaning that the ``peas in a pod'' phenomena \citep{weiss2018,wei2020,jiang2020,millholland2021} seen in single stars may be rare around binaries.

Compared to the  mechanism of small single planets being ejected around binaries because they migrate too slowly \citep{martinfitzmaurice2021}, our method of planet-planet resonant interaction leading to ejection is more efficient.

The transit, radial velocity, and microlensing detection methods have the capability to detect small circumbinary planets. Detection of the following types of circumbinary systems could validate our migration and resonant planet interactions method: evolved multi-planet systems in a resonant chain, multi-planet systems with a decrease in planet mass with increased semi-major axis, and young multi-planet systems with less massive inner planets and more massive outer planets not yet in a resonant chain. A significant increase in the number of known circumbinary systems will reveal the nature of the circumbinary planet distribution.

\section*{Acknowledgements}

We appreciate the comments from a referee that undoubtedly improved this paper. Hanno Rein and Dan Tamayo were great helps in assisting our implementation of stochastic forcing in \textsc{ReboundX}. We thank the planet groups at The Ohio State and Arizona, as well as Rosemary Mardling and Amaury Triaud, for fruitful discussions on this topic. Support for this work was provided by NASA through the NASA Hubble Fellowship grant HF2-51464 awarded by the Space Telescope Science Institute, which is operated by the Association of Universities for Research in Astronomy, Inc., for NASA, under contract NAS5-26555. This work was completed in part with resources provided by the University of Chicago Research Computing Center.

\section*{Data availability}
All simulation data is available on request by contacting the corresponding author

\bibliographystyle{mnras}
\bibliography{references}

\bsp	
\label{lastpage}
\end{document}